# Observation of Multiple Topological Corner States in Thermal Diffusion

Minghong Qi[1,2,3,4†], Yanxiang Wang[1,2,3,4†], Pei-Chao Cao[1,2,3,4], Xue-Feng Zhu[5], Fei Gao[1,2,3,4], Hongsheng Chen[1,2,3,4]*, and Ying Li[1,2,3,4]*

[1] *Interdisciplinary Center for Quantum Information, State Key Laboratory of Extreme Photonics and Instrumentation, ZJU-Hangzhou Global Scientific and Technological Innovation Center, Zhejiang University, Hangzhou 310027, China*

[2] *International Joint Innovation Center, The Electromagnetics Academy at Zhejiang University, Zhejiang University, Haining 314400, China*

[3] *Key Lab. of Advanced Micro/Nano Electronic Devices & Smart Systems of Zhejiang, Jinhua Institute of Zhejiang University, Zhejiang University, Jinhua 321099, China*

[4] *Shaoxing Institute of Zhejiang University, Zhejiang University, Shaoxing 312000, China*

[5] *School of Physics and Innovation Institute, Huazhong University of Science and Technology, Wuhan 430074, China*

† These authors contributed equally to this work.

*e-mail: hansomchen@zju.edu.cn; * e-mail: eleying@zju.edu.cn

**Higher-dimensional topological meta-materials have more flexible than one-dimensional topological materials, which are more convenient to apply and solve practical problems. However, in diffusion systems, higher-dimensional topological states have not been well studied. In this work, we experimentally realized the 2D topological structure based on a kagome lattice of thermal metamaterial. Due to the anti-Hermitian properties of the diffusion Hamiltonian, it has purely imaginary eigenvalues corresponding to the decay rate. By theoretical analysis and directly observing the decay rate of temperature through experiments, we present the various corner states in 2D topological diffusive system. Our work constitutes the first realization of multiple corner states with high decay rates in a pure diffusion system, which provides a new idea for the design of topological protected thermal metamaterial in the future.**

**Introduction**

Over the past few decades, the concept of topology has revolutionized the field of condensed matter physics[1][2], extending beyond just electronic band structures to a wide range of physical platforms, including photonic structures[3]-[8], acoustic[9]-[10] and mechanical[11]-[13] systems. In particular, topological insulators (TIs) have been extensively studied in the context of condensed matter physics and have now expanded to classical wave systems due to the similarity between wave equations and Schrödinger equations [14][15]. In higher dimensions, TIs exhibit more intriguing boundary states, such as high-order TIs (HOTIs)[16]-[19], Chern insulators[20]-[22], and topological semimetals[23]-[25]. Unlike conventional first-order TIs, HOTIs allow for topological states with a dimensionality that is two or more lower than the dimensionality of the system itself, supported at the boundaries of boundaries, such as corners. HOTIs have attracted significant attention in recent years due to their potential applications in topological lasers[26][27] and topological nanocavities[28][29].

While wave propagation has been the primary focus of previous topological research, the field of thermal metamaterials[30]-[32] has recently emerged as an exciting platform for exploring topological phenomena in diffusion systems. Thermal metamaterials have been designed to achieve new functionalities, such as shielding[33]-[36] and non-reciprocity[37]-[41], in flexible thermal regulation. Furthermore, due to the immune properties of topological states to defects and disorder, topological concepts have been introduced into thermal regulation[42][43], facilitating the study of robust edge states based on Su-Schrieffer-Heeger (SSH)[44]-[46] and non-Hermitian skin effects[47][48]. Very recently, diffusive topological edge states have been experimentally observed in a 2D SSH thermal lattice[49]. However, the majority of research has been limited to one-dimensional systems, hindering the discovery of unique diffusion topological states in higher dimensions. In addition, due to the limited research on topological states to the first few energy bands, high decay rate topological states have not yet been revealed.

In this work, we experimentally realized a two-dimensional topological thermal metamaterial based on the kagome lattice. Through theoretical analysis and directly observing the decay rate of temperature through experiments, we demonstrate that multiple high decay rate topological states exist between two types of merged cells in the pure diffusion system. We provide the first experiment realized the multiple corner states which have the superior local heat dissipation speeds. This breakthrough discovery provides a new idea for the design of the thermal metamaterial with topological protection in the future, with potential applications in thermal management, energy collection, and more.

**Thermal kagome lattice**

We present a 3D diffusive system that is based on the thermal analogue of 2D kagome lattices, featuring the density $\rho$, heat capacity $c_p$, and thermal conductivity $\kappa$ as shown in Figure 1a. Each unit cell comprises of three cylindrical sites with volume $V$. The cylinders are separated by a distance of $a/2$ while unit cells are separated by a distance of $a$ in $x$ direction and $\sqrt{3}a/2$ in $y$ direction. The interconnections between them are represented by the red (blue) rectangular rods with a cross-sectional area of $S_1$ ($S_2$), respectively. These rods correspond to the (nearest-neighbor) intra-cell ($D_1$) and inter-cell ($D_2$) couplings on the upward and downward-pointing triangles, respectively. (**Numerical simulations**). The cross-sectional area of the rods can be customized to control the heat exchange efficiency within and between the unit cells as shown in Eq. (1).

$$D_{1,2} = \frac{\kappa}{\rho c_p V a} S_{1,2} \tag{1}$$

We define the unit cell composed by three nodes specified by $j = 1, 2, 3$. Following the Fick's law, the temperature field evolution at each lattice site $T_{j,m}$ can be described as:

$$\begin{cases} -\partial_t T_{1,m} = D_1(T_{1,m} - T_{2,m}) + D_1(T_{1,m} - T_{3,m}) + D_2(T_{1,m} - T_{2,m+a_1}) + D_2(T_{1,m} - T_{3,m-a_2}) \\ -\partial_t T_{2,m} = D_1(T_{2,m} - T_{1,m}) + D_1(T_{2,m} - T_{3,m}) + D_2(T_{2,m} - T_{1,m-a_1}) + D_2(T_{2,m} - T_{3,m-a_3}) \\ -\partial_t T_{3,m} = D_1(T_{3,m} - T_{m,1}) + D_1(T_{3,m} - T_{2,m}) + D_2(T_{3,m} - T_{1,m+a_2}) + D_2(T_{3,m} - T_{2,m+a_3}) \end{cases} \tag{2}$$

where we define the translation direction vectors $\boldsymbol{a}_1 = \boldsymbol{x}$, $\boldsymbol{a}_2 = -\boldsymbol{x}/2 + \boldsymbol{y}$ and $\boldsymbol{a}_3 = \boldsymbol{x}/2 + \boldsymbol{y}$ with the same length $|\boldsymbol{a}| \equiv a$. Different from classical wave systems, the governing Hamiltonian of the thermal fields is anti-Hermitian $i\partial_t \boldsymbol{T} = H\boldsymbol{T}$, resulting in a purely imaginary eigen spectrum ($\omega$)[44]. Then we use the periodic boundary condition, the bulk Hamiltionian of thermal Kagome lattice $H$ can be written as :

$$H = -iD_2 \begin{pmatrix} 2\Delta+2 & -\Delta-e^{ik_xa} & -\Delta-e^{i(\frac{k_x}{2}-\frac{\sqrt{3}k_y}{2})a} \\ -\Delta-e^{-ik_xa} & 2\Delta+2 & -\Delta-e^{i(-\frac{k_x}{2}-\frac{\sqrt{3}k_y}{2})a} \\ -\Delta-e^{i(-\frac{k_x}{2}+\frac{\sqrt{3}k_y}{2})a} & -\Delta-e^{i(\frac{k_x}{2}+\frac{\sqrt{3}k_y}{2})a} & 2\Delta+2 \end{pmatrix} \quad (3)$$

where the ratio $\Delta = D_1/D_2$. The obtained imaginary eigenvalue corresponds to the thermal diffusion rate $\lambda$ of the lattice system that $\lambda = -\mathrm{Im}(\omega)$.

This tight-binding model is an extension of the 1D Su-Schrieffer-Heeger model, and its bulk topology can be characterized by the polarization[19], expressed as:

$$P = -\frac{1}{S}\iint A_i d^2k \quad (4)$$

where $d^2k$ is the area element in the Brillouin zone, $A_i = -i\langle u|\partial k_i|u\rangle$ is the Berry connection, with $i = x,y$ and $u$ being the Bloch functions, is the Berry connection of the lowest band, and S is the area of the first Brillouin zone. The polarization $(P_x, P_y)$ is identical to the Wannier center. Mirror symmetries constrain the Wannier center to two positions within each unit cell, which correspond to the two topologically distinct phases of the bulk. We refer to these as topologically "trivial" and "nontrivial" phases. Previous theoretical studies have shown that $(P_x, P_y)$ is entirely determined by the ratio $\Delta$. Note that although the values of $(P_x, P_y)$ depend on the choice of the unit cell, the Wannier center positions within the lattice are unambiguous. (See Supporting Information section 2).

To increase the bandgap, we reduce $S_1$ and increase $S_2$ to adjust $\Delta = 12.5$, and the bulk spectrum of the 2D thermal kagome model is shown in Figure 1b. It can be observed that the lower and middle energy bands have opened large bandgaps, as well as a flat band at the top of the middle band. To observe corner states in our design, we consider two types of unit cells, as shown in Figure 1c. The area inside the orange dashed triangle is composed of units with Δ, while the rest is composed of units with Δ' = 1/Δ.

The Hamiltonian is discrete for heat transfer in theory while the actual heat transfer is continuous, so the length of the connecting rods between elements in the thermal model is inaccurate, which can be corrected by multiplying a coefficient on the energy band (See Supporting Information section 3). The band structure of the system is shown in Figure 2a, revealing three edge states and three corner states in the $N$ modes. Notably, the states below the orange line are traditional states in the gap of the three energy bands discussed by Hamiltonians, while the colored states beyond the orange line are high decay rate topological states. Figure 2bc displays the temperature field distribution of two types of edge states located on the boundary between the two types of units or on the outer edge of the triangular region composed of extraordinary state units. For the three corner states, labeled I, II, and III, their temperature field distribution is shown in Figure 2d, where energy concentrates near the corner at the junction of the two types of units.

We perform finite element simulations using COMSOL Multiphysics, as shown in Figure 3. Taking 60 degrees above room temperature as the initial value of the heat, we attach the initial temperature to the cylinder unit according to the temperature distribution proportions of the 3 corner states in Figure 2d. At the same time, we select a cylindrical element that is in the bulk and apply the same initial heat, comparing the decay rate of the highest temperature under these conditions. Theoretically, the eigenvalue of the system should decay exponentially with time, with its decay rate corresponding to the imaginary part of the eigenvalue[44]. A larger decay rate will indicate a faster cooling ability. After 20 seconds, the highest temperature of corner state III has dropped to 315.9 K, while the highest temperature of corner state II is still 319.3 K. It is clear that the highest temperature of corner state III decays the fastest, followed by corner state II. The highest temperature decay rate of corner state I and bulk state is slower. To make the comparison clearer, we extend the time span to 120 seconds, and it is found that the bulk state has the slowest highest temperature decay rate, which is consistent with the prediction.

**Experiments**

To experimentally observe the corner states, we measured the maximum temperature of the heated spheres over time (**Experiments**). We used stainless steel powder 3D printing to design the system, and sprayed black paint on the surface to reduce its reflectivity and obtain accurate data from the thermal imager, as shown in Figure 4a. The cylindrical sites were heated and cooled using a small hot air gun and freezing spray to stimulate the bulk state and three corner states, as shown in Figure 3. (See Supporting Information section 4 for methods and experimental results with thermally insulated boundary conditions.)

According to theoretical predictions, the eigenvalue of the system should decay exponentially with time, and its decay rate, represented by $\lambda$, corresponds to the imaginary part of the eigenvalue. The decay rate of corner I (red line) in our experiments matched the predicted speed represented by $\lambda$ of corner state I (red dotted line), as shown in Figure 4b. Furthermore, it was larger than the decay rate of the bulk state (black line). These results suggest that the initial temperature field of a hot spot localized on the edge sphere of model I should be close to the edge state distribution in Figure 3, as confirmed by our measured results. Similarly, the decay rates for corner II (green line) and corner III (blue line) also matched the predicted speeds represented by their respective $\lambda$ (green/blue dotted line). The decay rates of corner II and III were significantly greater than that of corner I, which caused their maximum temperature to rapidly reduce to near room temperature, and caused their decay rates to gradually lose an exponential time dependence. The experimental results for the decay rates of the four cases were consistent with the simulation results, which can be found in Figure S2.

Experimentally measured thermal images of the temperature distributions are shown in Figure 4c. For the bulk state and corner I state, when the same initial temperature was applied, corner I had the lower maximum temperature with the most confined temperature field. This suggests that the initial hot spots of three corner states introduced less heating to the adjacent sites, demonstrating a well-localized heat diffusion. The larger decay rates of corner states II and III led to a shorter time for maintaining the temperature distributions of the eigenstates. Thus, the interception time was shorter (20 s) for corner II and III. Corner II had the lowest maximum temperature with the most confined temperature field. The experimental results for the four cases were consistent with the simulation results, which can be found in Figure 3.

**Conclusion**

We present theoretical explorations and experimental realizations of the 2D kagome thermal lattice with two opposite kinds of units splicing, extending higher-dimensional topological phases into purely diffusive heat-transfer systems. We have innovatively discussed high decay rate topological states and, for the first time, discovered and experimentally demonstrated multiple corner states. The high decay rate corner states exhibit higher diffusion rates than corner states in traditional bandgaps and have stable and robust distributions. These high decay rate topological phenomena in the diffusion system provide a new perspective for heat transfer control, which is expected to facilitate the design of novel thermal metamaterials and devices for efficient heat dissipation and thermal management with topological protection robustness. The proposed approach holds promise for developing a new class of thermal systems with enhanced thermal conductivity and unprecedented functionality, which could lead to new advances in a wide range of applications, including energy conversion, electronics cooling, and nanophotonics.

**Numerical simulations.** The parameters are: material parameter density $\rho = 8000$ kg/m$^3$, heat capacity $c_p = 500$ J/(kg·K) and thermal conductivity $\kappa = 16$ W/(m·K). We take 37 kagome units to form a regular hexagon with open boundary conditions and one triangle region was selected as the non-trivial part. Each kagome unit comprises three cylinders sites with a radius $R = 10$ mm and a height $H = 4$ mm. The intra-cell and inter-cell rods have same length with a unit length $a = 50$ mm. In the orange dashed triangle region, the cross-sectional areas of intra-cell ($S_1$) and inter-cell ($S_2$) rods are 1.6 mm$^2$ and 20 mm$^2$, respectively. Conversely, in the rest, the cross-sectional areas of intra-cell ($S_1$) and inter-cell ($S_2$) rods are 20 mm$^2$ and 1.6 mm$^2$, respectively.

**Experiments.** We measure and analyze the evolution of the temperature field under appropriate initial conditions. A stainless-steel sample was created using 3D metal printing, and black matte paint was sprayed on its surface to increase its reflection coefficient. The sample is placed on an adiabatic base made of thick styrofoam plate. To minimize the impact of the hot air gun on the surrounding units of the hot spot in the heating process (Figure 4a), a 10 cm x 10 cm rectangular hole is opened in the middle of the base. The room temperature is 293.15 K. The initial temperature of excited states is achieved using a small-caliber hot air gun and a freeze spray, as shown in Figure 2. The temperature of the hot air gun is over 380 K, ensuring that the sphere can be heated to 40 K above the room temperature. The freeze spray use HFC-134A as a cold source, ensuring that the sphere can be cooled to 40 K below the room temperature. The temperature fields are measured using an infrared camera Fotric 347. The maximum temperature of the heated sphere dropping to 312 K for the bulk state and corner state I is used to set $t = 0$. For corner states II and III, $t = 0$ is set when the maximum temperature of the heated sphere dropped to 303 K.

**Acknowledgments**

The work at Zhejiang University was sponsored by the National Natural Science Foundation of China (NNSFC) under Grants No. 92163123, No. 61625502, No.11961141010, and No. 61975176, the Top-Notch Young Talents Program of China, and the Fundamental Research Funds for the Central Universities.

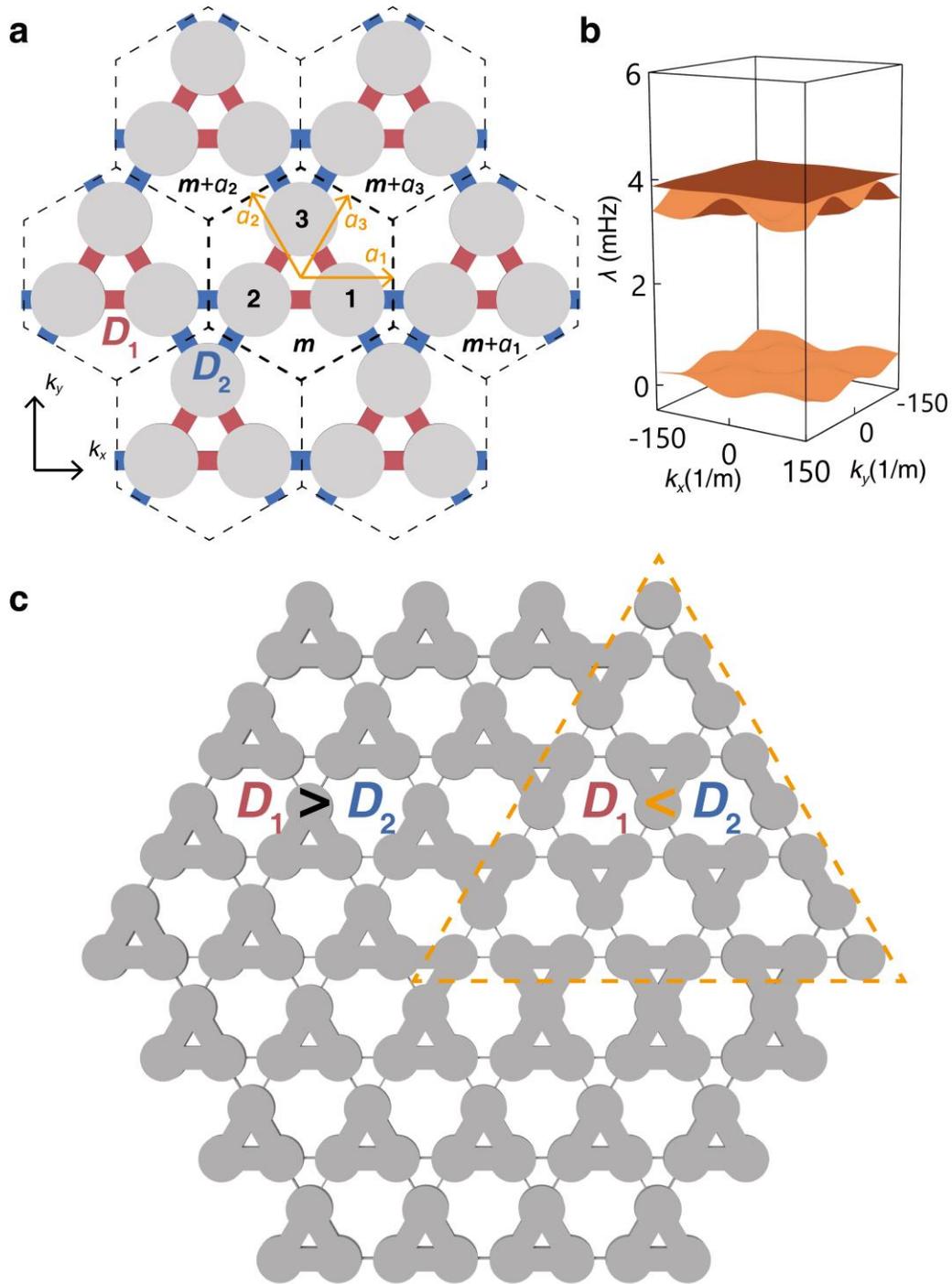

**Figure 1.** Heat diffusion in 2D thermal kagome lattices. a) 2D thermal kagome lattices composed of cylindrical sites and coupling rods. The intra-cell and inter-cell couplings are red and blue colored, respectively. b) The bulk spectrum of 2D thermal kagome model ($\Delta = 12.5$). c) Overview of the designed system with two types of the incorporated unit cells. The orange triangle represents a part where $D_1 < D_2$.

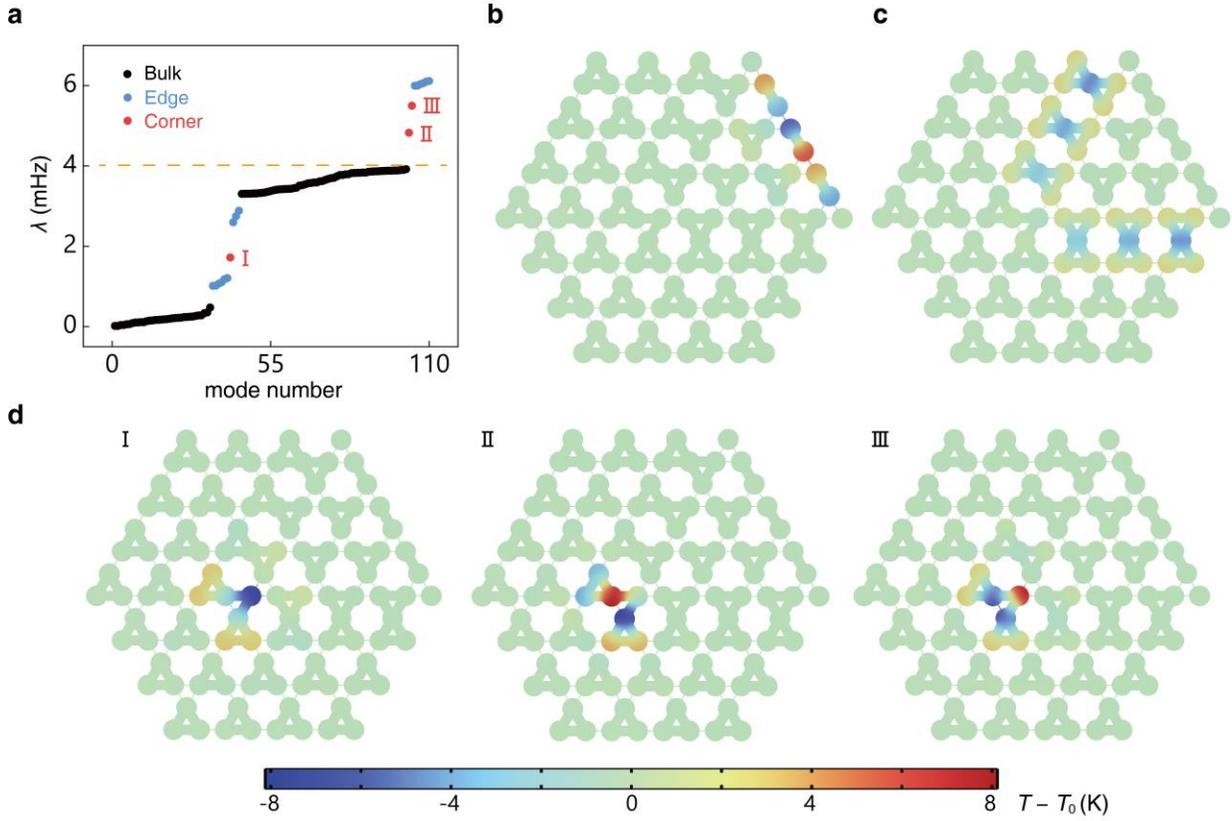

**Figure 2.** Edge states and corner states. a) The band structures of the system. The edge states are marked in blue. The three corner states are marked in red I, II and III. The colored states beyond the orange dashed line are high decay rate topological states. b,c) Temperature distributions of edge states with two different energy local positions. d) Temperature distributions of the three corner states.

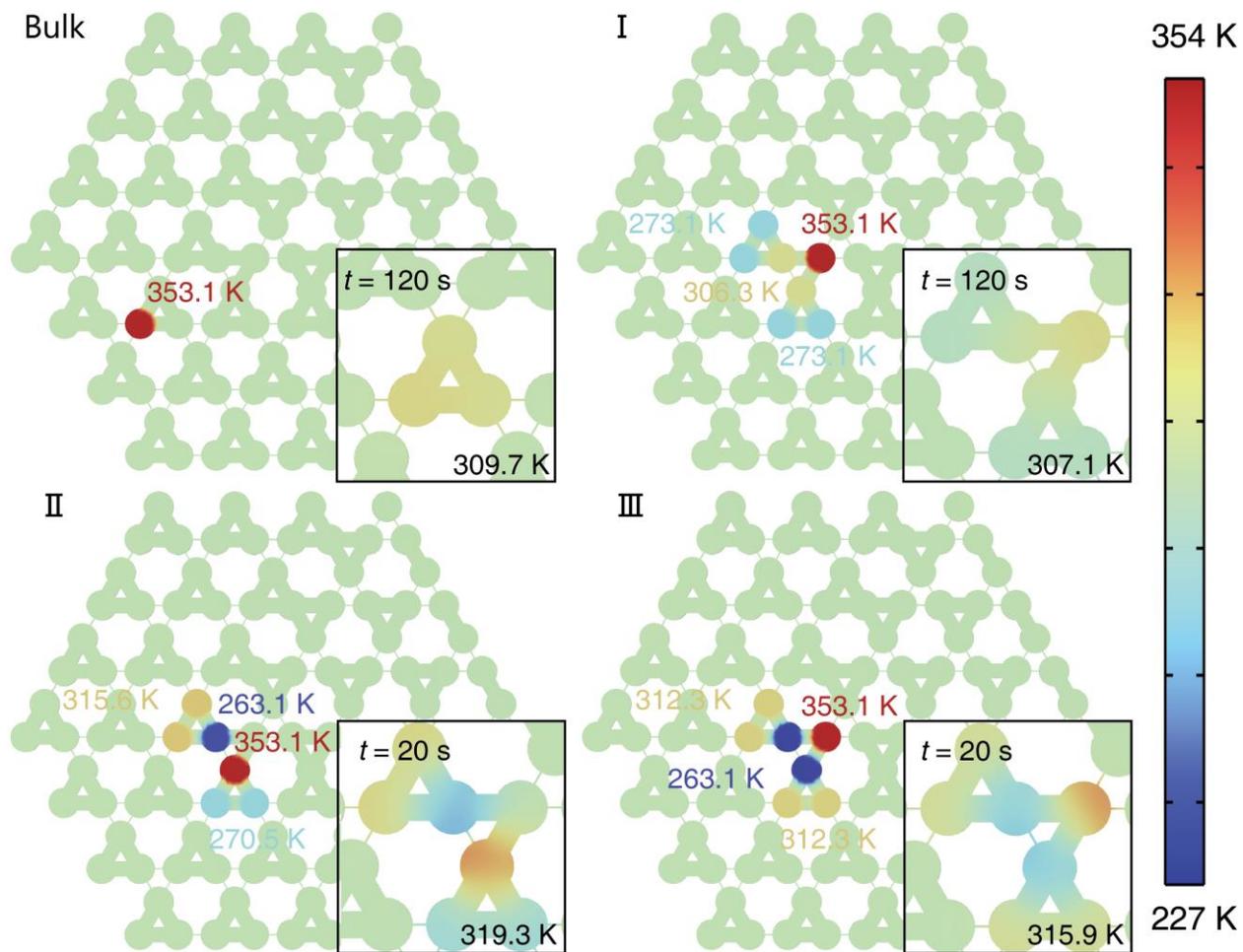

**Figure 3.** Simulation results. Temperature distribution diagram of bulk state and corner state I, II and III at $t = 0$ s. The black boxes are locally enlarged images of bulk state and corner state I at $t = 120$ s, corner state II and corner state III at $t = 20$ s. The black value in the lower right corner represents the highest temperature at this time.

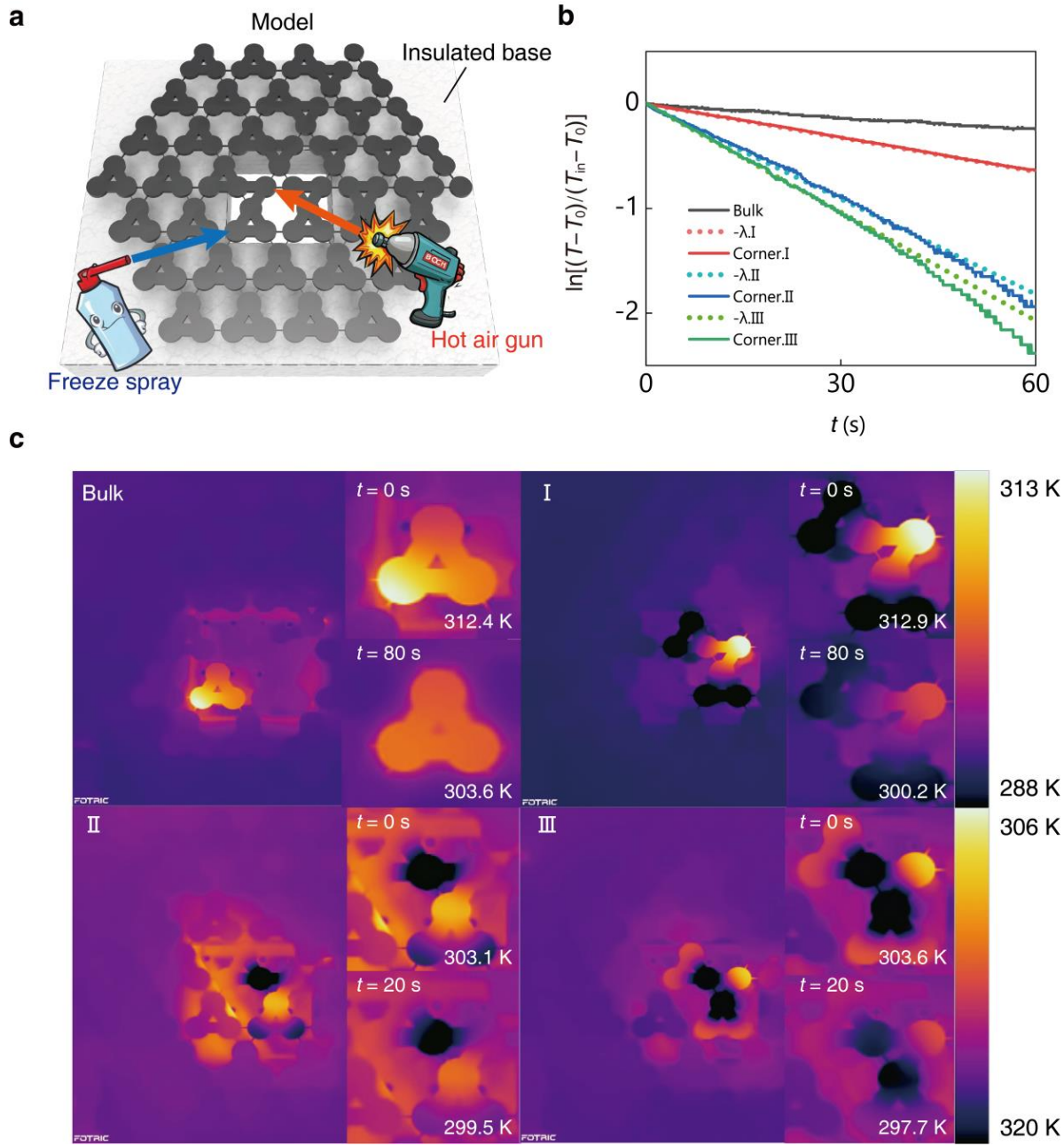

**Figure 4.** Experimental setup and results. a) Experimental Setup. The model is placed on an insulated base with a hole for heating and cooling. b) The maximum temperature decreasing rates. c) The overall and local enlarged thermal images. The values represent the highest temperature.